\documentclass[usenatbib]{mnras}

\RequirePackage[loading]{tracefnt}
\usepackage{graphicx}
\usepackage[space]{grffile}
\usepackage{latexsym}
\usepackage{textcomp}
\usepackage{longtable}
\usepackage{multirow,booktabs}
\usepackage{amsfonts,amsmath,amssymb}
\usepackage{natbib}
\usepackage{url}
\usepackage{hyperref}
\hypersetup{colorlinks=false,pdfborder={0 0 0}}
\newif\iflatexml\latexmlfalse
\usepackage[utf8]{inputenc}
\usepackage[ngerman,english]{babel}

\title[PSR J1946+3417: Mass and Formation]{A Massive Millisecond Pulsar in an Eccentric Binary}

\author[E.~D.~Barr et al.]
{\parbox{\textwidth}{E.~D.~Barr,$^{1}$\thanks{E-mail: \texttt{ebarr@mpifr-bonn.mpg.de}}
P.~C.~C. Freire,$^{1}$
M.~Kramer,$^{1}$
D.~J.~Champion,$^{1}$
M.~Berezina,$^{1}$
C.~G.~Bassa,$^{2}$
A.~G.~Lyne,$^{3}$
B.~W.~Stappers$^{3}$}\vspace{0.4cm}\\
\parbox{\textwidth}{
$^{1}$Max-Planck-Institut f\"ur Radioastronomie, Auf dem H\"ugel 69, 53121 Bonn, Germany\\
$^{2}$ASTRON, the Netherlands Institute for Radio Astronomy, Postbus 2, NL-7990 AA Dwingeloo, the Netherlands\\
$^{3}$Jodrell Bank Centre for Astrophysics, School of Physics and Astronomy, The University of Manchester, M13 9PL, UK}}

\begin{document}

\maketitle

\label{firstpage}

\begin{abstract}
The recent discovery of a population of eccentric ($e \, \sim \, 0.1$) millisecond pulsar (MSP) binaries with low-mass white dwarf companions in the Galactic field represents a challenge to evolutionary models that explain MSP formation as recycling: all such models predict that the orbits become highly circularised during a long period of accretion. The members of this new population exhibit remarkably similar properties (orbital periods, eccentricities, companion masses, spin periods) and several models have been put forward that suggest a common formation channel. In this work we present the results of an extensive timing campaign focusing on one member of this new population, PSR~J1946$+$3417. Through measurement of the both the advance of periastron and Shapiro delay for this system, we determine the mass of the pulsar, companion and the inclination of the orbit to be 1.828(22) $M_{\odot}$, 0.2656(19) $M_{\odot}$ and $76.4(6)^{\circ}$, under the assumption that general relativity is the true description of gravity. Notably, this is the third highest mass measured for any pulsar. Using these masses and the astrometric properties of PSR~J1946$+$3417 we examine three proposed formation channels for eccentric MSP binaries. While our results are consistent with eccentricity growth driven by a circumbinary disk or neutron star to strange star phase transition, we rule out rotationally delayed accretion-induced collapse as the mechanism responsible for the configuration of the PSR~J1946$+$3417 system.%
\end{abstract}

\section{Introduction}

Discovered in the Northern High Time Resolution Universe (HTRU-N) pulsar survey \citep{Barr_2013}, PSR J1946$+$3417 (hereon referred to as J1946) is a 3.17-ms spin period pulsar residing in an eccentric ($e \approx 0.13$) binary system with an intermediate mass companion ($M_{\rm min} \approx 0.26\ M_{\odot}$). 
At face value, the eccentricity of J1946's orbit appears to be at odds with the generally accepted model of millisecond pulsar (MSP) formation \citep{Alpar_1982,Bhattacharya_1991} in which the high spin period of the pulsar is attributed to the capture of angular momentum through mass transfer from a donor star. 
In this model, tidal forces during the long, steady period of accretion result in a strong circularisation of the orbit \citep[$e\approx 10^{-7}$--$10^{-3}$;][]{Phinney_1992,Phinney_1994}.

Until recently, fully-recycled MSPs with eccentric orbits were thought to be restricted to globular clusters (GCs), where the high stellar densities greatly increase the probability of $N$-body interactions. These interactions can act to decircularise the orbits of MSPs, inducing a change in the orbital eccentricity proportional to the mass of the encountering body and its closest approach \citep{Rasio_1995,Heggie_1996}. The discovery of PSR J1903$+$0327, a 2.15-ms MSP in orbit around a Solar-mass companion with $e \approx 0.44$ \citep{Champion_2008} challenged this model, suggesting that eccentric binary MSPs could also be formed in the Galactic field.
However the much lower stellar density of the Galactic field made it unlikely that PSR J1903$+$0327 had acquired its eccentricity from an encounter with a passing star.
Instead it has been proposed that the currently observed binary is the remnant of a dynamically unstable, hierarchical triple system in which the lowest-mass companion was ejected \citep{Freire_2011,Portegies_Zwart_2011,Pijloo_2012}. Credence has been lent to this theory in the form of the discovery of PSR J0337$+$1715, the first recycled pulsar in an extant stellar triple system \citep{Ransom_2014,Tauris_2014}.

Since the discovery of PSR J1903$+$0327 several additional eccentric binary MSPs have been found in the Galactic field. 
These include PSR J2234$+$0611 \citep{Deneva_2013,Antoniadis_2016_ArXiv}, PSR J1950$+$2414 \citep{Knispel_2015}, PSR~J0955$-$6150 \citep{Camilo_2015} and J1946 \citep[][ this work]{Barr_2013}. 
The four systems exhibit remarkably similar properties: binary periods, $P_{\rm b} = 22$ -- $32$ days, eccentricities, $e \approx 0.08$ -- $0.14$, minimum companion masses, $M_{\rm min}\approx0.19$ -- $0.25\ M_{\odot}$. 
Unlike PSR J1903$+$0327, with its solar-mass main sequence companion \citep{Champion_2008,Freire_2011}, these systems would appear to be classical MSP binaries, with binary periods and companion masses that follow approximately the $P_{\rm b}$-$M_{\rm WD}$ relation for MSPs with helium white dwarf (HeWD) companions \citep{Istrate_2014}. 
This suggests that these systems were formed through a different formation channel to PSR J1903$+$0327: a chaotic disruption of a triple system would tend to produce systems with widely different orbital characteristics. The same also applies to formation via a 3-body encounter in a globular cluster. The similarity in binary parameters suggests an orderly and repeatable formation common to all four MSPs.

Three hypotheses have been put forward to describe this formation channel: \begin{itemize}
\item \citet{Freire_2013} have posited that eccentric MSP-HeWD may be formed through the rotationally delayed, accretion-induced collapse (RD-AIC) of a super-Chandrasekhar mass white dwarf in a binary system. 
Here, the binary gains eccentricity through a combination of mass loss and kick during the supernova that results from the collapse. 
\item Alternatively, \citet{Antoniadis_2014} has posited that the observed eccentricity is a product of interaction between the binary and a circumbinary disk created during hydrogen flash episodes. 
These episodes can happen at the culmination of the long-term recycling phase and result in a super-Eddington mass-transfer rate, causing mass loss from the system. 
\item Finally, the system could have evolved as a normal MSP-WD system, but with the MSP later undergoing a phase transition in its core \citep{Jiang_2015}. 
\end{itemize}
\
These hypotheses and their predictions are discussed in detail in Section \ref{sec:discussion} in light of this work's results. 

This new population is also interesting for neutron star (NS) mass measurements, as the high eccentricities of the orbits result in large advances of periastron ($\dot{\omega}$). 
Under the assumption of general relativity (GR), measurement of this effect allows a measurement of the total mass of the system \citep{Taylor_1982}. 
Coupling measurements of $\dot{\omega}$ with measurement of other post-Keplerian parameters, such as Shapiro delay, can then allow a determination of the individual masses \citep[see e.g.][]{Freire_2011,Lynch_2012} and in a few cases provide a means by which gravity can be tested in the strong-field regime \citep[see e.g.][]{Kramer_2006}. 

Recent work has shown that MSPs have a broad range of possible masses, with accurately determined masses between 1.3655(21) $M_{\odot}$ \citep[PSR J1807$-$2500B;][]{Lynch_2012} and 2.01(4) $M_{\odot}$ \citep[PSR J0348$+$0432;][]{Antoniadis_2013}. However, it has been noted that the true distribution of MSP masses may be much wider than the above \citep{Freire_2008, van_Kerkwijk_2011} and may exhibit bimodality \citep{Antoniadis_2016_ArXiv_mass}. 

Accurate measurement of NS masses can allow for constraints to be placed on the NS equation of state (EoS). Here, the measurement of NS masses at the top end of the distribution rules out any EoS that predicts maximum NS masses lower than the measured value \citep[see e.g.][]{Demorest_2010}. Furthermore, in certain cases the measurement of MSP masses allows us to place model-dependent constraints on the birth mass of NSs \citep{Tauris_2011}, an important piece of the puzzle in understanding core-collapse physics. 
Finally, for the specific case of the aforementioned new class of millisecond pulsars in eccentric binaries, the mass measurements represent important tests of the hypotheses put forward for their formation.

Another reason why J1946 is interesting is because of its timing precision. A single 10-minute observation with Arecibo produces a TOA with rms uncertainty better than 300 ns. For this reason, this pulsar has been included in the pulsar timing arrays now being used to search for very low frequency gravitational waves \citep[e.g.][]{Hobbs_2010,Demorest_2012,Kramer_2013}. In this paper we will make use of data provided by these efforts. 

In this paper we present the results of a high-precision timing campaign on J1946. This campaign has provided both a precise mass measurement for J1946 and has allowed for direct testing of predictions associated with proposed formation mechanisms for eccentric MSP binaries. In Section \ref{sec:observations} we describe our observing setup and data taking, in Section \ref{sec:emission_properties} we discuss the J1946's emission properties, in Section \ref{sec:results} we present the results of our timing program, in Section \ref{sec:discussion} we discuss the implications of these results on proposed models for the formation of eccentric millisecond binary pulsars and finally in Section \ref{sec:conclusions} we present our conclusions.

\section{Observations}
\label{sec:observations}
After discovery, the pulsar was regularly followed up with the
Effelsberg and Lovell radio telescopes resulting in the
phase-coherent timing solution described in \citet{Barr_2013}.
These timing programs continue to this date with L-band observations at centre frequencies of 1.36 and 1.53 GHz and bandwidths of 400 and 200 MHz for the Lovell and Effelsberg telescopes, respectively. Observations with the Lovell are conducted with roughly weekly cadence with Effelsberg observations taken on a monthly basis. In this work we use all
TOAs from these two observing programmes taken until 2016 June 18.

In order to improve the timing precision and increase the
precision of measurement of relativistic effects, we 
observed J1946 with the L-wide receiver of the 305-m
Arecibo radio telescope 40 times over 2 years.
In these observations, we used the Puerto
Rico Ultimate Pulsar Processing Instrument (PUPPI, a clone of the
Green Bank Ultimate Pulsar Processing Instrument, GUPPI)\footnote{
  http://safe.nrao.edu/wiki/bin/view/CICADA/GUPPISupportGuide} as a
back-end. This allows simultaneous coherent dedispersion of the
 $\Delta f = 600$ MHz bandwidth provided by the receiver
(from 1130 to 1730 MHz) with a system temperature
$T_{\rm sys} = 30$ $\rm K$ and a gain $G = 10$ $\rm K/Jy$. 
Given the precise ephemeris derived from earlier Jodrell Bank and
Effelsberg data, the Arecibo observations could be made, from
the start, in coherent fold mode (with 512 channels, 2048 
phase bins and 4 Stokes parameters), which coherently
dedisperses and folds the data online, optimally removing the
dispersive effects of the interstellar medium.

The bulk of the observations were done under a program dedicated specifically to
this pulsar, where each data block includes 11 minutes of data. Here we obtained dense sampling of two full orbits:
June/July 2014 and April/March 2015. This is necessary for a precise measurement of the Shapiro delay;
the rate of observations was especially dense near superior conjunction, in
order to maximize sensitivity to this effect. The two dense campaigns
done almost one year apart also greatly improved our measurement of the
rate of advance of periastron, $\dot{\omega}$. The remaining Arecibo data were taken as part of the NANOGrav\footnote{\url{http://nanograv.org/}} project. This included both L-band and S-band (2.03 -- 2.73 GHz) observations.

The dedispersed pulse profiles obtained when averaging each data block
are then calibrated using the noise diode
observations that are taken with (almost) every single observation. 
The resultant calibrated pulse profiles are then cross-correlated
with a low-noise template using the procedure described in \citet{Taylor_1992} and implemented in
the \textsc{psrchive} software \citep{Hotan_2004,2012AR&T....9..237V}.

We then used \textsc{tempo}\footnote{\url{http://tempo.sourceforge.net/}}
to correct all TOAs using each telescope's clock corrections
and to convert them to the Solar System barycentre. To do this, the
motion of the radio telescopes relative to the Earth was calculated
using the data from the International Earth Rotation Service, and to
the barycentre using the DE421 Solar System
ephemeris\footnote{ftp://ssd.jpl.nasa.gov/pub/eph/planets/ioms/de421.iom.v1.pdf}.
Finally, the difference between the measured TOAs and those predicted
by a model of the spin and the orbit of the pulsar is minimized by
\textsc{tempo}, by varying the parameters in the model. To model the orbit, we used both the DD and DDGR models
\citep{1985AIHS...43..107D,1986AIHS...44..263D}. The DD
models the timing using a set of Post-Keplerian parameters, which
are derived from the timing data without any theoretical assumption.
The DDGR mode assumes the validity of GR
in the description of the orbital motion of the system and uses as
parameters the total mass of the system $M$ and the companion mass
$M_c$.

\begin{figure}
\begin{center}
\includegraphics[width=1\columnwidth]{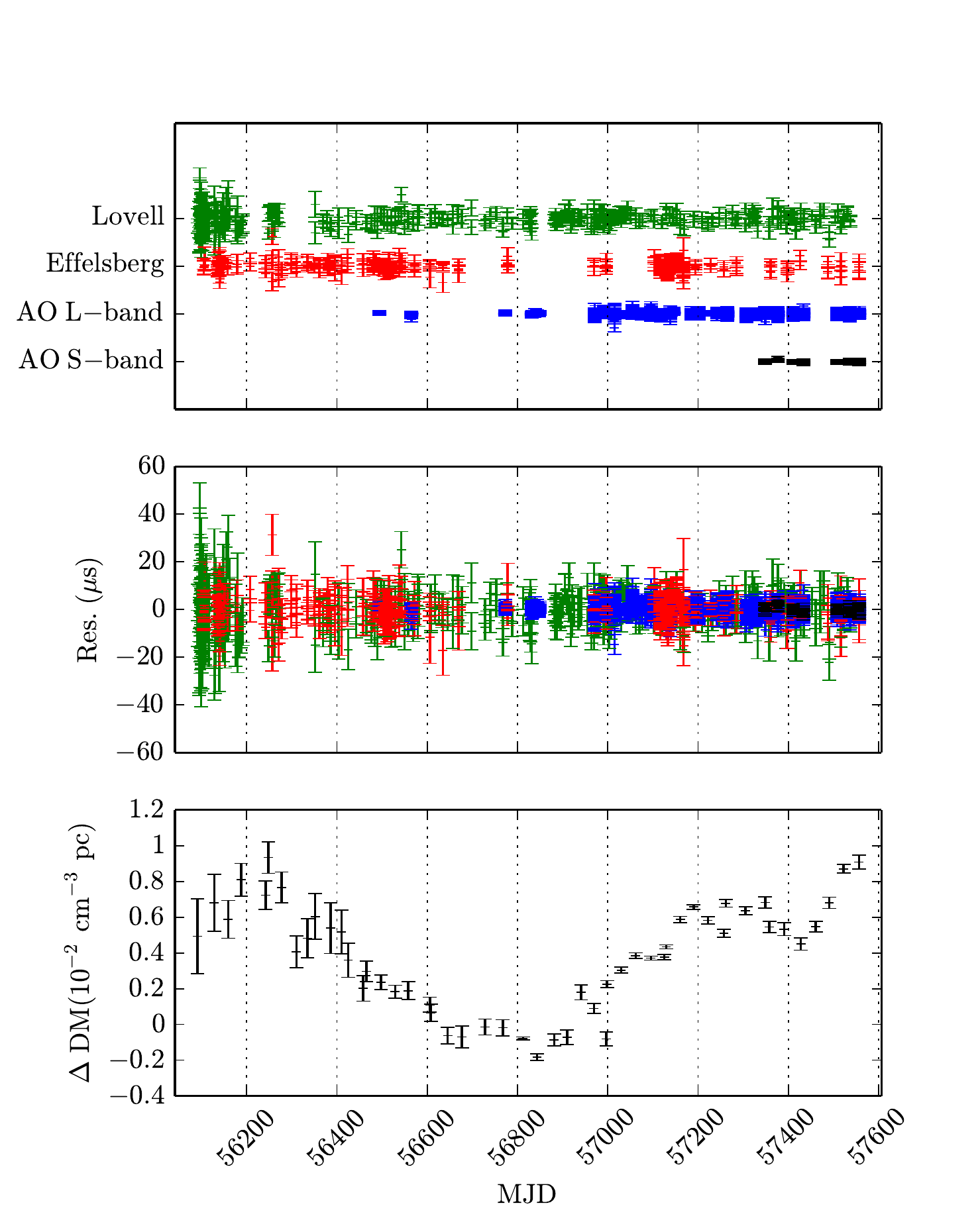}
\caption{{\label{fig:residuals}
{\em Top panel:} The temporal coverage of PSR~J1946$+$3417 observations. The residuals (TOAs minus DDGR model predictions) are split and offset to show the coverage with each of the Effelsberg, Lovell and Arecibo (both at L-band and S-band) telescopes. See the text for instrument descriptions. {\em Middle panel:} The same residuals as above now on a common set of axes. Shown are the $1$-$\sigma$ uncertainties. The residual root mean square is 1.3 $\mu$s. {\em Bottom panel:} The relative variation of DM at all observing epochs measured using L-band data from each telescope. These measurements were provided to \textsc{Tempo} during timing model fitting (see \ref{sec:dm_model}). Again the $1$-$\sigma$ uncertainties are shown.%
}}
\end{center}
\end{figure}

The residuals (TOAs minus model predictions) associated with the DDGR
model are displayed in Figure~\ref{fig:residuals}. There are some short term
trends in the residuals that point towards unmodeled systematics. 
The residual root mean square is 1.3 $\mu$s. 

\subsection{Emission properties}
\label{sec:emission_properties}

\begin{figure}
\begin{center}
\includegraphics[width=1\columnwidth]{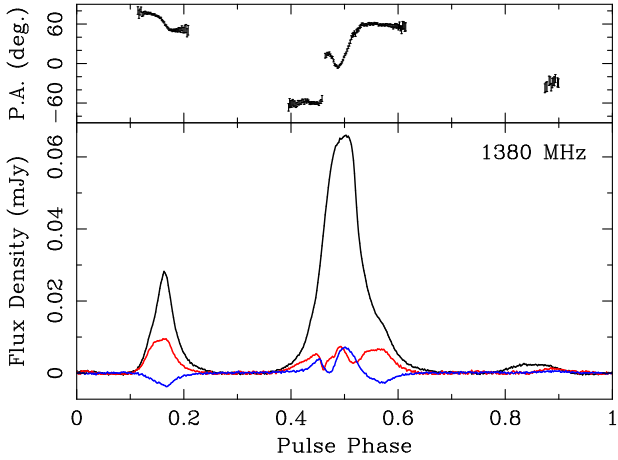}
\caption{{\label{fig:lband_prof}
The calibrated average L-band polarisation profile of PSR J1946$+$3417. The black line shows total intensity, the red line shows linear polarisation and the blue line shows circular polarisation. The upper plot shows the measured PA after correction for Faraday rotation assuming an RM of 4.6 rad m$^{-2}$. The three principle emission components are marked as ${a}$, ${b}$ and ${c}$.%
}}
\end{center}
\end{figure}

To examine the emission properties of J1946, all L-band data taken with the PUPPI instrument were cleaned of interference and flux and polarisation calibrated using noise diode observations taken prior to each integration.  Calibration was performed by the \textsc{pac} program (using the ``SingleAxis'' model) that is part of the \textsc{psrchive} software package \citep{Hotan_2004,2012AR&T....9..237V}.  Calibrated observations were then summed to form the high signal-to-noise ratio pulse profile shown in Figure \ref{fig:lband_prof}. To measure the position angle (PA) as a function of pulsar longitude, \textsc{psrchive} was used to perform rotation measure (RM) fitting. Using the \textsc{rmfit} program we find a best fit RM of 4.6$\pm$0.3 rad m$^{-2}$. Using this value, the PAs presented in Figure \ref{fig:lband_prof} have been corrected for Faraday rotation.

At L-band, J1946's profile exhibits three main emission components, named ${a}$, ${b}$ and ${c}$ in Figure \ref{fig:lband_prof} and located at $\sim80^{\circ}$, $\sim203^{\circ}$ and $\sim331^{\circ}$ in pulsar longitude. The distribution of emission regions across wide range of longitudes is indicative of a high degree of alignment between J1946's spin and magnetic axes. Details of the widths and polarisation fractions of each of these components is given in Table \ref{tab:components}. 

\begin{table} 
    \begin{tabular}{ c c c c c }
        \hline \hline
        Component & $W_{50}$ ($^\circ$) & $W_{10}$ ($^\circ$) & Linear ($\%$) & Circular ($\%$) \\ 
        \hline
         a & 14 & 34 & 43 & 13  \\ 
         b & 26 & 62 & 16 & 9  \\ 
         c & 34 & 50 & 37 & 8  \\
         \hline
    \end{tabular} 
    \caption{{The widths and polarisation fractions for components ${a}$, ${b}$ and ${c}$ as shown in Figure \ref{fig:lband_prof}. Here $W_{50}$ is the full width half maximum, $W_{10}$ is the full width at 10$\%$ maximum and ``Linear'' and ``Circular'' refer to the respective linear and circular polsarisation fractions. \label{tab:components}}}
\end{table}

\begin{figure}
\begin{center}
\includegraphics[width=1\columnwidth]{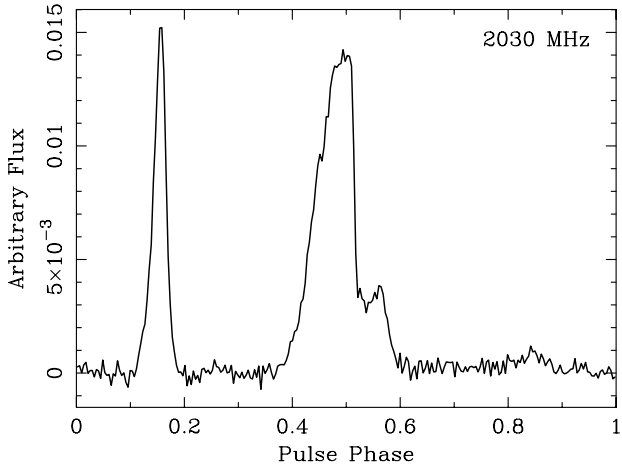}
\caption{{\label{fig:sband_prof}
The uncalibrated average S-band pulse profile of PSR J1946$+$3417. The profile has been rotated in phase such that its maximum aligns with that of the L-band profile.%
}}
\end{center}
\end{figure}

To investigate the evolution of J1946's profile with frequency several observations were taken using the Arecibo's S-Low receiver in combination with the PUPPI backend. Due to very strong RFI affecting these observations, it was not possible to calibrate these data. As such we present the uncalibrated, average S-band profile in Figure \ref{fig:sband_prof} merely to demonstrate the relative flux density changes between components. Here we see that component ${a}$ has brightened and the shoulder of component ${b}$ has developed into a distinct emission component. 

\begin{figure}
\begin{center}
\includegraphics[width=1\columnwidth]{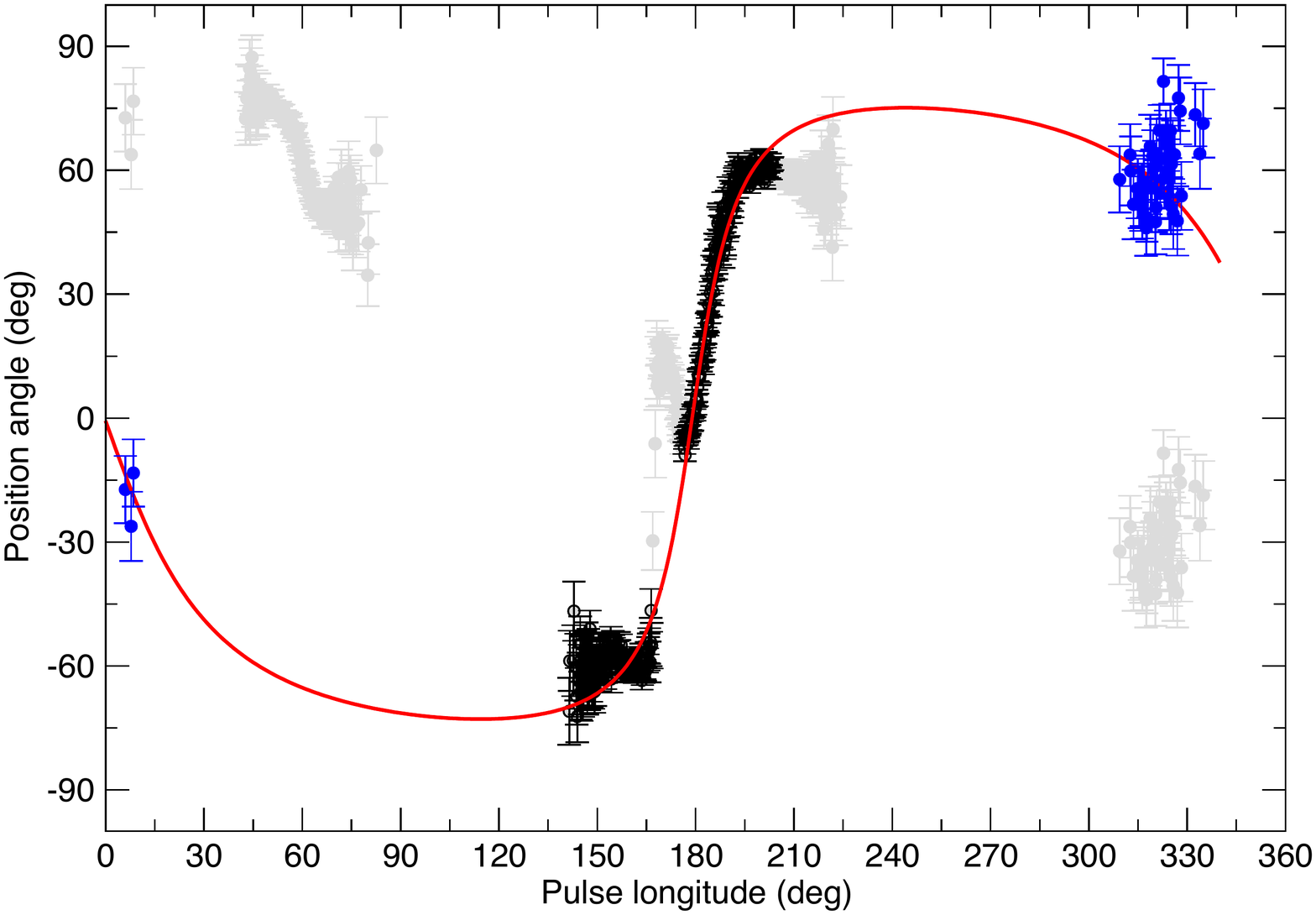}
\caption{{\label{fig:rvm_fit}
Modeling of PSR~J1946$+$3417's PA swing according to the RVM model. The red line shows a blind fit to the black (original) and blue (90 deg shifted) data. Grey data points were ignored during fitting. 
}}
\end{center}
\end{figure}

In Figure \ref{fig:rvm_fit} we model the observed PA swing according
to the Rotating Vector Model \citep[RVM;][]{Radhakrishnan_1969}. A fit
of the RVM to the data can yield the magnetic inclination angle,
$\alpha$, and the ``impact angle'', $\beta$, i.e. the angle between
the magnetic axis and the line-of-sight at the closest approach.  In
order for the pulsar to be observable, $\zeta = \alpha+\beta$ should
be comparable to the orbital inclination angle, $i$,
i.e.~$|\zeta-i|\le \rho$, where $\rho$ is the beam radius. While many
pulsars do not show regular RVM-like swings
\citep[e.g.][]{Johnston_2005}, millisecond pulsar PA swings are
especially difficult to be described by the RVM \citep[see
  e.g.][]{Xilouris_1998,Yan_2011}. Nevertheless, here we present a fit
to the data that is consistent with the orbital geometry determined
from pulsar timing.  The most notable feature is a distinct S-like
swing at longitudes from $\sim140$ deg to $\sim230$ deg. The centroid
of the swing shows a kink, very similar to that observed in
PSR~J1022$+$1001, which \citet{Ramachandran_2003} modeled as the
effect of return currents. Interestingly, also the step-like features
at the beginning and end of the S-swing show striking similarities
with the behaviour of PSR~J1022$+$1001.  However, in contrast to
PSR~J1022$+$1001, the pulsar has additional pulse components. Such a
feature is located about 180 deg away from the central S-swing, as
expected from emission from the other magnetic pole. Assuming that the
emission from this features shows an orthogonal jump relative to the
central swing, the data can be described by the same overall RVM swing
(see blue data points which were shifted by 90 deg from the observed
PAs).  In the shown fit, we ignored the kink and the trailing feature
of the central PA-swing (see grey data points). We also obviously
ignored a second S-like swing between 45 and 75 deg, corresponding to
the second strong pulse feature.  As has been previously speculated
\citep[see e.g.][]{handbook}, compact magnetospheres, as in the case
of the Crab pulsar or recycled pulsars, may encompass polar cap and
outer gap regions that may both be observable at radio frequencies. In
such a case, the latter would not be described by the same RVM and is
hence ignored here.  A blind fit to the black (original) and blue
(shifted by 90 deg) data leads the result shown in Figure
\ref{fig:rvm_fit} with $\alpha \sim 72$ deg and $\beta \sim 10$
deg. This is consistent with the orbital geometry determined from
pulsar timing. Given the relatively large pulse width for both main
and interpulse, it is also consistent with seeing emission from both
poles. It is remarkable that a RVM fit to a millisecond pulsar such as
J1946 produces a fit that is both plausible and consistent with the
timing results, under the following assumptions: a) two pulse features
correspond to two opposite magnetic poles, one showing an orthogonal
jump, while b) a second strong feature originates from a different
magnetospheric location such as an outer gap. The latter assumption
could be confirmed with a detection of gamma-ray emission, which would
then be expected to align with the radio component assumed to be from
an outer gap. Unfortunately, the spin-down luminosity and distance to
J1946 make such a detection unlikely \citep[see e.g.][]{Guillemot_2016}, 
and indeed, a search in \emph{Fermi} data did not reveal a corresponding source.

\subsection{DM Model}
\label{sec:dm_model}
To account for variations in J1946's DM, we used the ``DMX''
parameterization implemented in \textsc{tempo}. This calculates a DM
for consecutive intervals of time of uniform length, which in our case
we chose to be close to the orbital period, 27 days. The resulting DMs
are displayed in Figure \ref{fig:residuals}. These variations can be
attributed to turbulence in the ISM and their scale is consistent with
the measured variations for other MSPs with similar DMs \citep[see
  e.g.][]{Arzoumanian_2015}.  In this plot $\Delta DM \, = \, 0 \,\rm
cm^{-3} pc$ represents the assumed DM in Table 1. This value cannot be
determined precisely since there is substantial evolution of the pulse
profile in our band. This generates systematic non-zero TOA residuals
with observing frequency. These residuals can be be modeled adequately
with a single ``frequency derivative'' parameter, FD1 (see Table
\ref{tab:model}).

\section{Timing Results}
\label{sec:results}
Table \ref{tab:model} shows the parameters of coherent timing
solutions determined using the DD and DDGR models. The uncertainties
in this table represent 68.3 \% confidence levels. They were derived
from a Monte Carlo bootstrap method implemented in \textsc{Tempo}, we
made $2^{15}$ realizations of the TOAs, which vary around the observed
TOAs proportionally to their quoted errorbars (and with a distribution
similar to the observed residual distribution). For each realization,
a fit of the timing quantities is made and their statistical
properties are then analyzed, providing more conservative estimates of
the parameter uncertainties.

\begin{table*}
\begin{minipage}{\textwidth} 
\begin{center}
\label{tab:model}
\begin{tabular}{lrr}
\hline \hline
Time scale & TDB & TDB \\
Solar System ephemeris & DE421 & DE421 \\
Binary model & DD & DDGR\\
Reference epoch (MJD) & 56819.452908 & 56819.452908 \\
\hline
Timing parameters & & \\
\hline
Right ascension, $\alpha$ (J2000.0) & 19$^{\mathrm h}$46$^{\mathrm m}$25\fs131145(20) & 19$^{\mathrm h}$46$^{\mathrm m}$25\fs131145(20) \\
Declination, $\delta$ (J2000.0) & $+$34\degr17\arcmin14\farcs68898(14) & $+$34\degr17\arcmin14\farcs68898(14) \\
$\mu_\alpha$ (mas yr$^{-1}$) & $-$7.01(23) & $-$7.03(22) \\
$\mu_\delta$ (mas yr$^{-1}$) & $+$4.51(21) & $+$4.51(20) \\
Spin frequency, $\nu$ (Hz) & 315.443563698779(8) & 315.443563698779(8)\\
First derivative of $\nu$, $\dot{\nu}$ (10$^{-16}$ Hz s$^{-1}$) & $-$3.130(3) & $-$3.130(3)\\
DM (cm$^{-3}$ pc)\footnote{This is not an absolute DM measurement as the absolute DM is degenerate with the observed frequency-dependent profile evolution. Instead we here present the reference DM value ($\Delta {\rm DM} = 0$) of the model presented in Figure \ref{fig:residuals}.} & 110.209335 & 110.209335\\
Rotation measure (rad m$^{-2}$) & +4.6(3)& +4.6(3) \\
Profile frequency dependency parameter, FD1 & 0.000054(3) & 0.000054(3)\\
Orbital period, $P_{\rm b}$ (days) & 27.01994783(5) & 27.01994783(5) \\
Projected semi-major axis, $x$ (lt-s) & 13.8690718(5) & 13.8690724(22)  \\
Orbital eccentricity, $e$ & 0.134495389(17) & 0.134495376(10)  \\
Epoch of periastron, $T_0$ (MJD) & 56819.4529124(10) & 56819.4529118(5)  \\
Longitude of periastron, $\omega$ (\degr) & 223.364186(14) & 223.364177(14) \\
Advance of periastron, $\dot{\omega}$ (\degr yr$^{-1}$) & 0.001363(9) & \\
\hline
Derived parameters & & \\
\hline
Total mass of binary, $M_{\mathrm{t}}$ (M$_{\odot}$) & 2.094(22) & 2.094(22)\\
Companion mass, m$_2$ (M$_{\odot})$ & 0.31(5) & 0.2656(20) \\
$s\, \equiv \, \sin i$ & 0.964(8) & \\ 
Galactic longitude, $l$ & 69.29 & 69.29 \\
Galactic latitude, $b$ & 4.71 & 4.71 \\ 
Distance, $d$ (kpc) & 5.1 & 5.1\\
$\mu_{\rm tot}$ (mas yr$^{-1}$) & 8.33(22) & 8.35(21)\\
Proper motion position angle, Equatorial, $\phi_\mu$ (\degr) & 302.7(1.4) & 302.7(1.4) \\
Proper motion position angle, Galactic, $\phi_\mu$ (\degr) & 2.4(1.4)  & 2.3(1.4)\\
Spin period, $P$ (ms) & 3.17013917885772(8) & 3.17013917885772(8) \\
First derivative of spin period, $\dot{P}$ (10$^{-21}$) & 3.145(3) & 3.145(3) \\
Quadratic Doppler shift, $\beta$ (10$^{-23}$s$^{-1}$) & 1.35(4) &  1.36(3)\\
Intrinsic $\dot{P}$, $\dot P_{\rm int}$ (10$^{-21}$s$^{-1}$) & 3.132(3) & 3.132(3) \\
Characteristic age, $\tau_{\rm c}$ (Gyr) & 16.0 & 16.0 \\
Transverse magnetic field at the poles, $B_0$ (10$^{7}$ Gauss) & 9.5 & 9.5 \\
Rate of rotational energy loss, $\dot{E}$ (10$^{33}$ erg s$^{-1}$) & 3.9 & 3.9\\
Mass function, $f$ (M$_{\odot}$) & 0.0039233396(4) & 0.0039233401(18)\\
Mass ratio, $q \equiv M_{\mathrm{p}}/M_{\mathrm{c}}$ & 5.8 & 6.9 \\
Orbital inclination, $i$ (\degr) &  74.7(1.6)  &  \\
Pulsar mass, m$_{\rm p}$ (M$_{\odot}$) & 1.78(5) & 1.828(22) \\
Residual r.m.s., ($\mu$s) & 1.3 & 1.3 \\
\hline
\end{tabular}
\end{center}
\end{minipage}
\caption{{
\label{tab:model}
PSR J1946$+$3417 ephemerides assuming the DD and DDGR binary models (see Section \ref{sec:observations}). The DMX model was used to account for long-term dispersion measure variations in the pulsar. Both ephemerides were created from TOAs taken with the Effelsberg and Arecibo telescopes over the course of three years. Numbers in parentheses represent the 1-$\sigma$ uncertainties in the trailing digit as determined by the Monte-Carlo bootstrap method implemented in \textsc{tempo}.  The characteristic age, spin-down luminosity and surface magnetic field strengths were calculated using the Shklovskii-corrected period derivative. The position, frequency and DM are all measured with respect to the given reference epoch. The total mass is a fitted parameter in the DDGR ephemeris, but a derived parameter in the DD ephemeris. The parameters $\dot{\omega}$, $s$ and $i$ are not fit for in the DDGR model.}}
\end{table*}

\subsection{Masses}

The pulsar's orbital eccentricity allows a detection of the
advance of periastron, $\dot{\omega}$. If both components are compact,
i.e. we here assume a HeWD companion, then this effect depends only on the Keplerian orbital
parameters, which are already known precisely, and the total mass of the binary $M$.
Thus $M$ can be derived from a measurement of $\dot{\omega}_{\rm GR} $ \citep{Taylor_1982}:
\begin{equation}
\label{eqn:M}
M = \frac{1}{T_{\odot}} \left[ \frac{\dot{\omega}_{\rm GR}}{3} (1- e^2) \right]^{\frac{3}{2}} \left(\frac{P_{\mathrm{b}}}{2\pi} \right)^{\frac{5}{2}}.
\end{equation}
where $T_{\odot} = G M_{\odot} c^{-3} = 4.925490947 \mu \rm s$ is a
solar mass ($M_{\odot}$) in time units ($c$ is the speed of light and
$G$ is Newton's gravitational constant).
This yields, according to the DDGR model, $M \, =\, 2.094(22)\, M_{\odot}$.

\begin{figure*}
\begin{center}
\includegraphics[width=2\columnwidth]{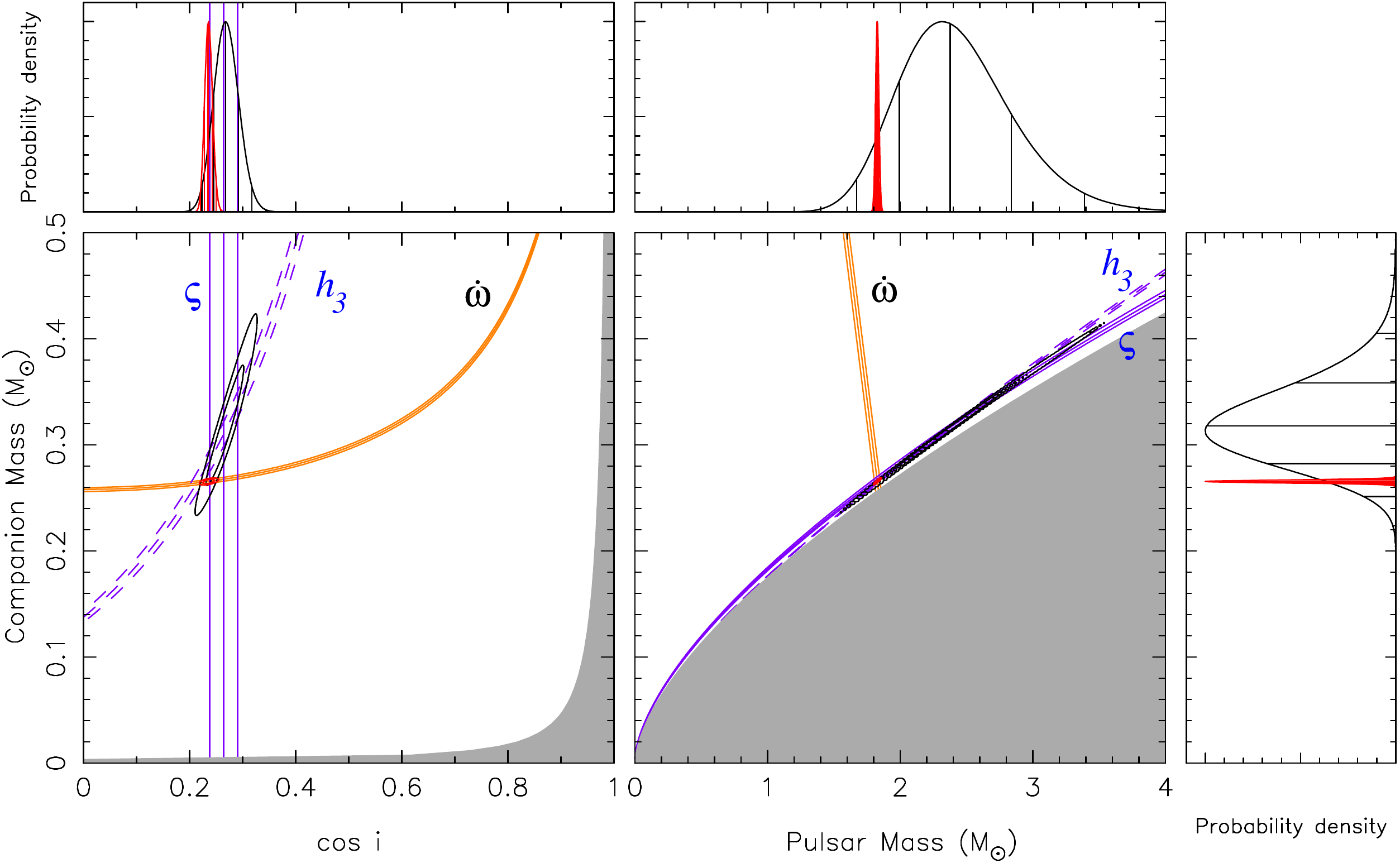}
\caption{{\label{fig:massmass}
Current constraints from timing of PSR~J1946$+$3417. The orange lines correspond to the nominal and  $\pm 1$-$\sigma$ uncertainties the rate of advance of periastron $\dot{\omega}$, and the purple lines correspond to the nominal and  $\pm 1$-$\sigma$ uncertainties associated with the two orthometric Shapiro delay parameters, $\varsigma$ (solid) and $h_3$ (dashed). These values and uncertainties are as derived in Table 1.  The black contour levels contain 68.27 and 95.45\% of the 2-D probability density function (pdf) derived from the quality of the timing solution at each point of the $M_c$ - $\cos i$ plane using only the Shapiro delay, the red contour lines the pdf obtained from the Shapiro delay plus the assumption that the $\dot{\omega}$ is due only to the effects of GR. {\em Left}: $M_c$ - $\cos i$ plane. The gray area is excluded by the physical constraint $M_p \, > \, 0$. {\em Right}: $M_c$ - $M_p$   plane. The gray region is excluded by the mathematical constraint $\sin i \leq 1$. {\em Top and right panels}: Pdfs for $\cos i$, $M_p$ and (on the right) $M_c$, derived from marginalizing the 2-D pdfs in the main panel for these quantities. When $\dot{\omega}$ is taken into account (red), the precision of the mass estimates improves by a factor of about 50.
}}
\end{center}
\end{figure*}

We also measure the Shapiro delay. Using the DD parameterization, we
obtain $s\, \equiv \, \sin i \, = \, 0.964(8)$ and $m_2 \, = \, 0.31(5) \, M_{\odot}$.
This would result in a pulsar mass $m_p \, = \, 2.3(6)  \, M_{\odot}$,
a measurement that indicates a large NS mass but at a low precision. This measurement
of the Shapiro delay is nevertheless highly significant.
Using the orthometric parameterisation \citet{Freire_2010} we obtain
$\varsigma \, = \, 0.763(22)$ and $h_3 \, =\, 0.68(5)\, \rm \mu s$
(the latter is estimated not from a direct fit, but from the uncertainty of
$m_2$ when $\sin i$ fixed at its best value, using the bootstrap Monte Carlo method described above).
The $h_3$ parameter is measured with a significance of 14 $\sigma$, implying a very
robust detection of the effect.
The 1-$\sigma$ constraints introduced by these parameters are presented in
Fig.~\ref{fig:massmass} as the purple lines. Although all the available timing data were used in the fitting described here, it is worth noting the importance of the Arecibo data taken around superior conjuction as without these data the Shapiro delay parameters do not have significant measurements.

In order to verify the masses derived from the Shapiro delay, we 
implemented a Bayesian scheme similar to that described in \citet{Splaver_2002}.
We sampled the
quality of the fit for a wide region in the $M_c - \cos i$ plane,
depicted in the left panel of Figure~\ref{fig:massmass}.  For each
grid point in this region, we calculate the Shapiro delay parameters assuming
GR to be the correct theory of gravity in this regime and then fix
them in the timing solution. We then fit for all
other timing parameters. The quality of the fit is quantified by the
residual $\chi^2$ of the resulting solution, the lower this is the better the
fit. From the resultant $\chi^2$ grid we derive a 2-D probability distribution 
function (pdf) using the Bayesian specification in \citet{Splaver_2002}.
This is then converted to a similar 2-D pdf in the $M_c - M_p$ plane
(right panel) using the mass function relation. 
The black contours of both panels
include 68.27 and 95.45\% of the total probability of each pdf.
We then marginalize the 2-D pdfs to derive 1-D pdfs for $M_c$, $\cos i$
and $M_p$, the latter are presented in the
top and right marginal panels in black. These pdfs yield, for their medians and
$\pm 68.3 \%$ percentiles the following values: $M_c\, = \, 0.317^{+0.041}_{-0.035}\, M_{\odot}$,
$M_p \, = \, 2.37^{+0.46}_{-0.38}\, M_{\odot}$ and $i \, = \, 74.4\, \pm \, 1.4^\circ$.

Since we measure a third Post-Keplerian parameter ($\dot{\omega}$),
the mass equations are over-determined and we have a test of the self-consistency
of general relativity. The theory passes the test because the $\dot{\omega}$
line crosses the region where $\varsigma$ and $h_3$ intersect. However, given
the large uncertainty in the prediction of the total mass from the Shapiro delay
only, this test has very low precision.

Although this Shapiro delay measurement is not very impressive {\em per se},
in combination with the total mass constraint it allows very precise
mass measurements.
If for each point in the $M_c - \cos i$ plane we fix
the Shapiro delay parameters {\em and} fix the value of $\dot{\omega}$
to the GR prediction for the total system mass at that point
(and then continue as described above), we derive a much narrower 2-D pdf.
The latter is illustrated in
Figure~\ref{fig:massmass} by the red contour lines, which include
68.27\% and 95.4\% of its total probability. 
Marginalizing these pdfs along the different axes we 
obtain much narrower 1-D pdfs for $\cos i$, $M_p$ and $M_c$,
which are also depicted in red in the marginal panels.
Their medians and 1-$\sigma$ percentiles occur at
$M_{p} \, = \, 1.827(13) \, M_{\odot}$, $M_{c} \, = \, 0.2654(13) \, M_{\odot}$
and $i = 76.41^{+0.44}_{-0.42}\, ^\circ$, respectively. This is
consistent with the masses and uncertainties in Table 1 (which were derived
from the bootstrap Monte Carlo method), but
with slightly smaller error bars. For this reason, we will from now
on use the more conservative values and uncertainties in Table 1.

This whole exercise has, nevertheless, allowed us to understand why are
the mass estimates produced by DDGR so precise compared to those
produced by the DD model. 
\\

\subsection{Transverse velocity}
\label{sec:transverse}
From Table \ref{tab:model} we see that J1947 exhibits a total proper motion of 8.35(21) mas yr$^{-1}$. Using the NE2001 Galactic free electron density model \citep{Cordes2002}, we find a DM-derived distance to J1946 of $5\pm2$ kpc, assuming a 40\% uncertainty \citep[see][for discussion on NE2001 model uncertainties]{Deller_2009}. This corresponds to a total transverse velocity of $200\pm60$ km s$^{-1}$, a typical value for recycled pulsars \citep{Hobbs_2005}. The motion of the pulsar is directed out of the Galactic plane, with a Galactic position angle of $(2.3\, \pm \, 1.4)^{\circ}$; this corresponds to an equatorial position angle of $(302.7\, \pm\, 1.4)^{\circ}$.

\subsection{Optical detectability}

\citet{Antoniadis_2016_ArXiv} recently announced the optical detection of the HeWD in the eccentric MSP binary system PSR\,J2234$+$0611. In  the case of J1946, archival SDSS\,DR7 \citep{Abazajian_2009,Adelman-McCarthy_2009} and IPHAS DR2 \citep{Barentsen_2014} imaging of the J1946 field show no counterpart at the location of the pulsar. Because of the proximity of a brighter star ($r^\prime=14.6$) at $2\farcs5$ from the pulsar position, we conservatively estimate an upper limit of $r^\prime>20$. At the DM-derived distance of 5.1\,kpc, the \citet{Green_2014,Green_2015} model of reddening $E_{B-V}$ as a function of distance along the line-of-sight towards J1946 predicts $E_{B-V}=0.28$. With the extinction coefficients of \citet{Schlafly_2011}, this translates to an SDSS $r^\prime$-band absorption of $A_{r^\prime}=0.640$ and hence $(m-M)_{r^\prime}=14.17$. Our conservative upper limit on the $r^\prime$ brightness of the companion to J1946 only rules out the hottest and youngest white dwarf models by \citet{Serenelli_2001}. Even extending the upper limit to $r^\prime>25$ would only rule out models with $T_\mathrm{eff}>9200$\,K and $\tau_\mathrm{WD}<0.4$\,Gyr. Hence, we conclude that detection of the white dwarf companion to J1946 is unlikely.

\section{Discussion}
\label{sec:discussion}
\subsection{Model comparisons}
Here we review the implications of PSR J1946's measured parameters on three proposed models of eccentric MSP formation. 

\subsubsection{Circumbinary disk}

\citet{Antoniadis_2014} has proposed that eccentricity could be induced in the orbits of MSPs through dynamical interaction of the binary system with a circumbinary disk. This model makes three predictions that PSR J1946 can be used to test, namely:

\begin{enumerate}
\item The companions of eccentric MSPs should be HeWDs that follow the $P_{\rm b}$--$M_{\rm WD}$ relation for LMXB evolution \citep{Tauris_1999}.
\item Both the masses and systemic velocities of eccentric binary MSPs should closely resemble those of circular binary MSPs.
\item There is an upper limit to the amount of eccentricity that can be induced by interaction with a circumbinary disk, $e \lesssim 0.15$.
\end{enumerate}

The $P_{\rm b}$--$M_{\rm WD}$ relation of \citet{Tauris_1999} can be represented by
\begin{equation}
\frac{M_{\rm WD}}{M_\odot} = \left(\frac{P_{\rm b}}{b}\right)^{1/a} + c,
\end{equation}
where $a$, $b$ and $c$ are constants that depend on the chemical composition of the donor star. For Pop I. stars, $a=4.5$, $b=1.2\times10^5$ and $c=0.12$. Using these values, we obtain a prediction for the mass of J1946's companion of 0.27 $M_\odot$, in good agreement with our measured mass of 0.265(1) $M_\odot$.

As discussed in Section \ref{sec:transverse}, J1946 has a transverse velocity of $200\pm60$ km s$^{-1}$. Although on the high side, this is consistent with the known distribution of transverse velocities for recycled pulsars \citep{Hobbs_2005}. Similarly, while J1946 is relatively massive at 1.83(1) $M_\odot$, its mass is consistent with the known distribution of MSP masses, which is known to range from 1.3655(21) $M_{\odot}$ \citep[PSR J1807$-$2500B;][]{Lynch_2012} to 2.01(4) $M_{\odot}$ \citep[PSR J0348$+$0432;][]{Antoniadis_2013}.

Together, given a measured eccentricity of $e\simeq0.13$, we find that J1946 satisfies all three testable predictions of the \citet{Antoniadis_2014} circumbinary disk model.

\subsubsection{RD-AIC}

\citet{Freire_2013} have proposed that eccentric MSP-HeWD systems may be formed through the rotationally delayed, accretion-induced collapse (RD-AIC) of a super-Chandrasekhar mass white dwarf in a binary system.

The RD-AIC hypothesis makes several testable predictions:

\begin{enumerate}
\item HeWD companions in the RD-AIC scenario should have masses similar to the  $P_{\rm b}$--$M_{\rm WD}$ relation for LMXB evolution \citep{Tauris_1999}, but slightly lower (because of the slight orbital expansion during AIC) and orbital periods of between 10 and 60 days.
\item The resultant MSP in the RD-AIC scenario should have a mass in the range 1.22-1.31 $M_\odot$.
\item The post-AIC binary system should have a peculiar space velocity (and in particular a vertical velocity) $\lesssim10$ km s$^{-1}$.
\end{enumerate}

While PSR J1946's companion mass satisfies prediction (i), the high pulsar mass is clearly not compatible with prediction (ii).
Regarding prediction (iii), that is also incompatible with the observed proper motion. It is in principle possible that a pulsar with small peculiar velocity might display a significant proper motion (due to Galactic rotation), however, if that is the case, the latter should be confined to the Galactic plane. Instead, the large proper motion we observe is almost perpendicular to the plane. Since the pulsar is located close to the Galactic plane, this observed perpendicular velocity is therefore its vertical velocity, which is indeed far to large for the prediction of \citet{Freire_2013}. We can therefore rule out RD-AIC as being the mechanism responsible for PSR J1946's observed properties.

\subsubsection{Strange Star Scenario}

Another formation mechanism worth discussing in light of J1946's measurements, is that of transition of a massive NS into a strange quark star (SS). \citet{Jiang_2015} have proposed that a sufficiently massive NS may undergo phase transition into an SS if its core density is above the critical threshold for quark deconfinement. Baryonic mass converted to binding energy during this transition induces an eccentricity in the system. For phase transitions that occur towards the end of, or after, the LMXB phase, the resultant orbit will not re-circularise, leaving an eccentric MSP binary.

The NS-SS phase transition model has three key predictions:

\begin{enumerate}
\item The masses of SS MSPs can lie in a wide range depending on the amount of mass that is transferred into the binding energy.
\item The companions of SS MSPs should be HeWDs with a narrow range of possible masses ($M_{\rm WD} \sim 0.25$-$0.28$ $M_\odot$).
\item Having likely undergone a conventional supernova to produce the initial NS, eccentric SS MSP binaries should have kinematic properties similar to those of the known binary MSP population.
\end{enumerate}

It is clear that with only one measured mass prediction (i) is not particularly constraining. However, we may invert this prediction and ask what amount of mass would be required to be lost from the system to induce the observed eccentricity. Assuming no kick during the quark nova, we can relate the observed eccentricity to the total mass of the system via $e = \Delta M / (M_p + M_c)$ \citep{Bhattacharya_1991}. Here we calculate a total gravitational mass loss of 0.28 $M_\odot$, suggesting a pre-transition NS mass of $\sim 2.1~M_\odot$, larger than any measured NS mass to date.  

As noted in our discussion of the \citet{Antoniadis_2014} model, the observed kinematic properties of J1946 are consistent with those of the known MSP population, satisfying prediction (iii). Similarly, our companion mass measurement further satisfies prediction (ii). 

Due to the similarity of the predictions made by both \citet{Antoniadis_2014} and \citet{Jiang_2015}, we cannot use J1946 on its own to differentiate between the NS-SS or circumbinary disk model of eccentric binary MSP formation. However, if we assume that the fraction of mass converted to binding energy is similar for all quark novae, that there is a common NS EoS and that the critical density for quark deconfinement is common for a given EoS, then we expect there to be a relatively small range of viable SS masses for eccentric MSPs. With this in mind, the measurement of a $1.39(1)~M_\odot$ pulsar in another eccentric binary MSP system (PSR~J2234$+$0611; Stovall et al., in prep.) poses problems for the NS-SS transition model. \citet{Antoniadis_2016_ArXiv} determine the pre-transition mass of PSR~J2234$+$0611 to be $\sim1.6~M_\odot$. Given that the pre-transition mass of PSR~J2234$+$0611 is lower than the post-transition mass of J1946, we find it highly unlikely that these systems share a NS-SS formation channel and, by extension, that NS-SS transition is the common formation channel for eccentric MSP binaries.

\section{Conclusions}
\label{sec:conclusions}

In this paper, we have presented an overview of a four-year, multi-telescope timing campaign on the eccentric millisecond binary pulsar, PSR~J1946$+$3417. This has campaign has led to precise mass measurements of the component masses in the system and the orbital inclination ($M_{p} \, = \, 1.828(22) \, M_{\odot}$, $M_{c} \, = \, 0.2656(19)\, M_{\odot}$ and $i = 76.4(6)^{\circ}$) under the assumption of GR.
The mass of the companion is similar to that expected from the \citet{Tauris_1999} prediction, while the pulsar mass is the third largest measured to date. This indicates that massive neutron stars are not rare in the Universe: indeed, out of 35 neutron stars with precisely measured masses \citep{2016arXiv160302698O},
PSR~J1946$+$3417 and three others -- J1614$-$2230 \citep{Fonseca_2016}, J0348+0432 \citep{Antoniadis_2013} and J1012+5307, \citep{Antoniadis_2016_ArXiv_mass} -- have masses larger than $1.8 \, M_{\odot}$, i.e., these massive NSs constitute $\sim \, 10$\% of the known population.
The component masses and the astrometric properties of the system derived here allowed a of test several proposed mechanisms for the formation of eccentric MSP binaries in the Galactic field such as PSR~J1946$+$3417. The large pulsar mass and large vertical velocity ($200\pm60$ km s$^{-1}$) of the system are incompatible with the rotationally delayed accretion-induced collapse hypothesis \citep[RD-AIC;][]{Freire_2013}, effectively ruling it out as the mechanism responsible for the formation of PSR~J1946$+$3417 and the other systems in its class. We find that our results satisfy the predictions of formation via either dynamical interaction of a circular MSP binary with a circumbinary disk \citep{Antoniadis_2014} or transition of a massive neutron star into a strange quark star \citep{Jiang_2015}. However, we note that in combination with the recent results of \citet{Antoniadis_2016_ArXiv}, our results strongly disfavour the model of \citet{Jiang_2015} as being responsible for the observed properties of eccentric binary MSP systems.

The hypothesis proposed by \citet{Antoniadis_2014} for the formation of PSR~J1946$+$3417 and similar eccentric systems is the only one the three discussed here that is not excluded by current measurements. However we note that the degree to which circumbinary disks can raise orbital eccentricity is currently a matter of debate \citep[see e.g.][]{2016arXiv160505752R}. Further theoretical work on the role of a circumbinary disk is needed, along with more sources in this class, to determine if this is the evolutionary channel by which these objects are formed.

\section*{Acknowledgements}
The Arecibo Observatory is operated by SRI International under a cooperative agreement with the National Science Foundation (AST-1100968), and in alliance with Ana G. M\selectlanguage{ngerman}é\selectlanguage{english}ndez-Universidad Metropolitana, and the Universities Space Research Association. This work was partly based on observations with the 100-m telescope of the MPIfR (Max-Planck-Institut f\selectlanguage{ngerman}ü\selectlanguage{english}r Radioastronomie) at Effelsberg. Pulsar research at JBCA and access to the Lovell Telescope is supported by a Consolidated Grant from the UK Science and Technology Facilities Council (STFC).  P. F. gratefully acknowledges financial support by
the European Research Council for the ERC Starting grant BEACON under contract No. 279702.

\bibliographystyle{mnras}
\bibliography{bibliography/converted_to_latex.bib%
}

\end{document}